\documentclass[a4paper, 12pt]{article}
\begin{document}
\title{Quantum statistics and multiple particle production
\thanks{Supported in part by the KBN grant 2P03B 093 22}
}
\author{K.Zalewski\\
M.Smoluchowski Institute of Physics,
 Jagellonian University,\\ Cracow, Reymonta 4, 30 059 Poland,\\ e-mail:
zalewski@th.if.uj.edu.pl \\ and \\ Institute of Nuclear Physics
PAN, Cracow
 }
\maketitle
\begin{abstract}
Effects of quantum statistics are clearly seen in the final states of
high-energy multiparticle production processes. These effects are being widely
used to obtain information about the regions where the final state hadrons are
produced. Here we briefly present and discuss the assumptions underlying most
of these analyses.
\end{abstract}
Bose-Einstein correlations.\vspace{1cm}

\section{Introduction}

Differences between quantum and classical (Boltzmann) statistics show up in a
variety of systems. Here we will discuss systems, where the energy per particle
is of the order of 100 MeV or more. In such systems, produced in high energy
scattering processes, correlations due to quantum statistics are clearly seen
in the data: Identical bosons seem to attract each other, identical fermions
seem to repel each other. The quantitative description of these effects is
interesting both for their own sake and in order to disentangle other more
subtle correlations. There is still another motivation for this study, however,
which is somewhat controversial, but very stimulating.

Much work has been done on quantum statistics in multiple particle production
processes. For reviews see e.g. \cite{WIH} and \cite{CSO}. According to most of
this work, the study of interparticle correlations due to quantum statistics
yields valuable information about interaction regions, i.e. about the regions
where the hadrons are produced. The interaction regions are difficult to study,
because they are both small and short-lived. Their typical sizes are of the
order of fermis and also their life-times are of the order of fermis. A fermi
in time is the time necessary to cross the distance of one fermi at the speed
of light i.e. about $3 \times 10^{-24}$ sek. Very few methods for studying
interactions regions are available.

An added attraction is that the predictions from the recognized theories for
the new data from the heavy ion collider RHIC turned out to be completely wrong
(references can be traced e.g. from \cite{ZAL}). The name coined for this
disaster is \textit{the RHIC puzzle}. Consequently, both the general
theoretical framework and the specific phenomenological assumptions have to be
reanalyzed. In this paper we will present the basic assumptions of the most
popular models.

In low energy scattering, i.e. when the center-of-mass kinetic energy of the
two colliding particles is of the order of $1$ GeV, when several particle are
produced their distribution is roughly spherically symmetric. The first model
to gain wide popularity, Fermi's model, used a slightly modified microcanonical
distribution. For each multiplicity of particles, each final state with total
energy and total momentum equal to their initial values was assumed to be
equally probable. At higher energies spherical symmetry breaks down and the
model does not work any more. At low energies, however, it seemed good until
paper \cite{GOG} got published.

In this paper the authors studied the opening angles i.e. the angles between
the momenta for pairs of pions. From Coulomb interactions one would
qualitatively expect that for pairs of like-sign charged pions the opening
angles would be on the average smaller than for pairs of unlike-sign charged
pions. Quantitatively, however, one can easily estimate that this effect is
rather small. The authors expected, nevertheless, a similar effect, because it
had been predicted that in the $\pi^+\pi^-$ system there is a resonance, now
known as the $\rho$ meson, while for like sign pairs no resonance had been
expected. To their surprise, the experimental result was just the opposite. The
opening angles  for pairs of like-sign pions tended to be smaller. The result
got explained in the seminal paper \cite{GGL} as a result of the Bose-Einstein
statistics of the identical pions. For many years the effect had been known as
the GGLP effect. We will review the GGLP paper in the following section, but
now let us discuss high energy scattering.

We will consider central heavy ion (e.g. gold-gold) scattering at
center-of-mass energies of the order of $100$ GeV per pair of colliding
nucleons. In such collisions, in the center-of-mass system, due to Lorentz
contraction, both nuclei take the form of thin pancakes. When the two pancakes
fly through each other there are many nucleon-nucleon interactions, but at this
energies the directions of flight of the nucleons change little and the
pancakes survive. When they fly away from each other, many strings are formed
and stretch connecting the color charges in one nucleon with the opposite color
charges in the other. These string are a characteristic feature of quantum
chromodynamics. When two opposite electric charges interact, the well-known
field of forces extends over all space. In quantum chromodynamics the
corresponding field is confined to a thin (diameter of the order of one fm)
tube with the opposite color charges at the two ends. In a high energy central
heavy ion collision many strings are produced and exist simultaneously. It is
plausible that they merge and produce a well-defined, roughly cylindrical
region with the two pancakes at its ends. This region is presumably first
filled with quarks, antiquarks and gluons. Only later hadrons, mostly pions,
emerge from it.

There is a number of questions one would like to ask. What is the size and
shape of this region? What is its life-time and what is the duration of
hadronization? Note that the life-time and the duration of hadronization are in
general different. For instance, the region could live for $8$ fm without
emitting hadrons and then emit all the hadrons within 2 fm. If the content of
the region can be considered a phase, it would be interesting to know its
equation of state. This would be valuable information for cosmological models
of the early stages of the Universe expansion and for models of the interiors
of neutron stars -- perhaps some of them have quark-gluon cores. Another set of
questions is about the transition of this stuff into hadrons: is it a phase
transition or a cross-over? If it is a phase transition, is it first order or
continuous? If it is first order, what is the latent heat?

\section{GGLP or HBT or BE correlations}

Let us start this section with a remark about terminology. Starting from the
seventies, the original acronym GGLP was being gradually replaced by the
acronym HBT in honor of Hanbury-Brown and Twiss, who several years before the
GGLP paper \cite{GGL} had  made a related discussion in astronomy \cite{HBT}.
They used successfully the Bose-Einstein correlations among photons to measure
the diameters of distant stars. Now the name GGLP is hardly ever used. Kopylov
and Podgoretskii explained \cite{KOP} how, starting from a more general
formulation, one can obtain the GGLP case by making a parameter tend to zero
and the HBT case by making the same parameter tend to infinity. Thus, GGLP and
HBT are two different limiting cases. In fact, some people replace HBT by the
more neutral BEC standing for Bose-Einstein correlations.

In order to facilitate comparison with subsequent developments, we will present
the GGLP results using a more general notation. GGLP start with an input single
particle density operator -- not to be confused with the actual single particle
density operator for the system  -- in the form

\begin{equation}\label{}
  \hat{\rho}_I = \int d^3x |\textbf{x} \rangle \rho(\textbf{x})\langle\textbf{ x}|.
\end{equation}
This corresponds to particles being produced incoherently from point sources.
Function $\rho(\textbf{x})$ gives the distribution of the sources in space. In
this model the full information about the size and shape of the interaction
region is given when function $\rho(\textbf{x})$ is known.  The corresponding
input single particle density matrix reads

\begin{equation}\label{}
  \rho_I(\textbf{p};\textbf{p}') = \int \rho(\textbf{x})e^{i\textbf{qx}},
\end{equation}
where $\textbf{q} = \textbf{p} - \textbf{p}'$. Note that given the density
matrix $\rho_I(\textbf{p};\textbf{p}')$ one can unambiguously obtain the
distribution of sources $\rho(\textbf{x})$ just by inverting the Fourier
transformation. Unfortunately, this nice feature of the theory will be lost,
when the theory is made more realistic.

Interpreting the input density matrix as a density matrix, one would obtain the
single particle momentum distribution

\begin{equation}\label{}
\Omega_1(\textbf{p}) = \rho_I(\textbf{p};\textbf{p}) = \int
d^3x\rho(\textbf{x}) = 1.
\end{equation}
This is obviously unrealistic. E.g. very large momenta are forbidden by energy
conservation. GGLP, however, interpreted $\Omega_1(\textbf{p})$ as a weight for
the states allowed by energy and momentum conservation. Thus, the result just
means that the single particle distributions should be calculated from Fermi's
model.

For pairs of identical particles, in general,

\begin{equation}\label{}
  \Omega_2(\textbf{p}_1,\textbf{p}_2) \neq \rho_I(\textbf{p}_1;\textbf{p}_1)\rho_I(\textbf{p}_2;\textbf{p}_2),
\end{equation}
because the right-hand side does not have the right symmetry with respect to
the exchange of $\textbf{p}_1$ and $\textbf{p}_2$. GGLP symmetrized it to get

\begin{equation}\label{}
\Omega_2(\textbf{p}_1,\textbf{p}_2) =
\rho_I(\textbf{p}_1;\textbf{p}_1)\rho_I(\textbf{p}_2;\textbf{p}_2)+
\rho_I(\textbf{p}_1;\textbf{p}_2)\rho_I(\textbf{p}_2;\textbf{p}_1).
\end{equation}
For instance, substituting for the density distribution of sources
\begin{equation}\label{gaussi}
  \rho(\textbf{x}) = \frac{1}{\sqrt{2\pi R^2}^3}e^{-\frac{\textbf{x}^2}{2R^2}}
\end{equation}
one finds

\begin{equation}\label{}
\Omega_2(\textbf{p}_1,\textbf{p}_2) = 1 + e^{-R^2\textbf{q}^2}.
\end{equation}
Again, as a distribution of momenta this is untenable, but interpreted as a
weight it nicely reproduces the enhancement at small momentum differences seen
in the data.

The GGLP paper has been very influential. It still is the most quoted paper in
the field. It has, however, its weak points. On the technical side, it is easy
to calculate the weights, but then the integration over momentum space is
needed. In the usual case, when neither the nonrelativistic approximation, nor
the ultrarelativistic approximation (all masses tending to zero) is justified,
this is a cumbersome task. In practice, as far as I know, no one got with this
approach beyond three identical particles, while in heavy ion collisions, at
high energies hundreds of identical particles are being produced. Moreover, in
order to calculate the momentum integral, one needs the exact numbers of all
the other particles produced and this is usually not available. A reasonable
approximate solution of this problem has been found and will be presented in
the next section.

 Also the physics behind the model is doubtful. Time does not appear explicitly.
 As easily checked, this corresponds to the assumption that all the identical
 particles are produced instantly and simultaneously at some time $t_0$. This
 is a most unlikely scenario. Just reflect at the question, in which reference
 frame this assumption should be satisfied. Moreover, classically one would like
 to find the probability distribution for a particle being produced at point
 $\textbf{x}$ with momentum $\textbf{p}$. In order to get consistency with
 quantum mechanics one has to compromise. In the GGLP model the particle is
 produced at point $\textbf{x}$ and, in agreement with quantum mechanics, its momentum
 can be anything with equal probability. This is then modified by rejecting the
 states forbidden by energy and/or momentum conservation. It is much more
 likely, however, that the particles are produced as wave packets with some finite
 variances of position and momentum.

 \section{Two particle (reduced) correlation functions}

 Consider a two-particle correlation function

\begin{equation}\label{}
  C_2(\textbf{p}_1,\textbf{p}_2) =
  \frac{\Omega_2(\textbf{p}_1,\textbf{p}_2)}{\Omega_{2bckg}(\textbf{p}_1,\textbf{p}_2)}.
\end{equation}
Here $\Omega_2$ is the experimental two-body momentum distribution and
$\Omega_{2bckg}$ is the corresponding distribution with the Bose-Einstein
correlations switched off. The latter cannot be obtained from the data without
further assumptions, but the experimental groups have various methods for
getting reasonable  approximations to it. It would be more in agreement with
the terminology used in statistics to put into the denominator the product of
single particle distributions instead, but the correlation function defined
here is more convenient to study separately the correlations due to quantum
statistics. Assuming that this correlation function can be calculated from the
GGLP approach without phase space integrations, one finds

\begin{equation}\label{kopfor}
  \Omega_2(\textbf{p}_1,\textbf{p}_2) = \Omega_{2bckg}(\textbf{p}_1,\textbf{p}_2)\left(1 +
  \frac{|\rho(\textbf{p}_1,\textbf{p}_2)|^2}{\rho(\textbf{p}_1,\textbf{p}_1)\rho(\textbf{p}_2,\textbf{p}_2)}\right).
\end{equation}
From now on we skip the subscript $I$, though $\rho$, strictly speaking, is not
quite the single particle density matrix of the system. Formula (\ref{kopfor}),
which has been proposed by Kopylov \cite{KOP1}, is plausible and can be easily
used for comparison of models with experiment. One measures $\Omega_2$ and
divides it by the estimated $\Omega_{2bckg}$ to obtain the experimental $C_2$.
This is compared with the $C_2$ calculated from the model. The procedure is
approximate, but it avoids the integration over momentum space. Thus, the
information how many and what kinds of particles have been produced is not
needed.

For instance, for the Gaussian input density (\ref{gaussi}) the calculated
correlation function is

\begin{equation}\label{}
  C_2(\textbf{p}_1,\textbf{p}_2) = 1 + e^{-R^2\textbf{q}^2}.
\end{equation}
Comparing it with the data one obtains the root-mean-square radius of the
interaction region $R$. Here by assumption this region is spherically
symmetric. A natural generalization \cite{GRA}, \cite{BER} is to replace the
product $R^2q^2$ in the exponent by $q_L^2R_L^2 + q_o^2R_o^2 + q_s^2R_s^2$,
where $q_L$, $q_o$ and $q_s$ are, respectively, the components of $\textbf{q}$
along the beam axis and along some two axes perpendicular to it.  Since the
maximum length of the strings, achieved just before they break, increases with
energy, one would expect for central collisions at high energies $R_L \gg R_o
\approx R_s$. Experimentally it is observed that $R_L$ is comparable with $R_o$
and $R_s$. This has been explained as follows \cite{BOW}. The correlations due
to Bose-Einstein statistics are visible only for values of $|\textbf{q}|$ of
the order of $R^{-1}$ or less. Therefore, they can be used only to measure the
size of the region where particles with similar momenta are produced. This
region, named by Sinyukov the homogeneity region, is in general smaller than
the total interaction region.

The correction of the physical assumptions is much harder and will be discussed
in the following sections. Essentially there are two strategies: the method of
wave packets, with its close relative the method of covariant currents, and the
methods inspired by the concept of the Wigner function. This leaves aside the
approaches where a complete model is proposed, which can be used to calculate
anything. In particular it may be used to calculate the correlations among the
identical particles at small $|\textbf{q}|$ whether or not they have something
to do with some interaction region. For an example in this category cf. e.g.
\cite{ANH}.

\section{Wave packets and covariant currents}

Let us replace the GGLP single particle input density operator by

\begin{equation}\label{}
  \hat{\rho} = \int d^4x_s \int d^4p_s |\psi_{x_sp_s}\rangle
  \rho(x_s,p_s)\langle \psi_{x_sp_s}|,
\end{equation}
which corresponds to the single particle input density matrix

\begin{equation}\label{}
  \rho(\textbf{p};\textbf{p}') = \int d^4x_s \int d^4p_s \psi_{x_sp_s}(\textbf{p})
  \rho(x_s,p_s) \psi_{x_sp_s}(\textbf{p}').
\end{equation}
Here the time dependence of the density matrix is not explicitly written.
Function $\rho(x_s,p_s)$ is the distribution of the space-time four-vectors and
the energy-momentum four-vectors defining the sources. The distribution of
particles, however, is consistent with quantum mechanics because of the wave
functions $\psi_{x_sp_s}$. There are various ways of using this scenario.

Kopylov and Podgoretskii, who introduced it \cite{KOP2},  assumed that the
sources differ only by their positions in space-time. Thus

\begin{equation}\label{}
  \psi_{x_sp_s}(\textbf{p}) = e^{ipx_s}\phi(\textbf{p}),
\end{equation}
where the fourth component of $p$, necessary to calculate the product $px_s$,
is given by $p_0 = \sqrt{\textbf{p}^2 + m^2}$ and thus, is not an independent
variable. The density matrix is

\begin{equation}\label{}
  \rho(\textbf{p};\textbf{p}') = \phi(\textbf{p})\phi^*(\textbf{p}') \int d^4x_s \int d^4p_s
  \rho(x_s,p_s)e^{iqx_s}
\end{equation}
and

\begin{equation}\label{}
\Omega_1(\textbf{p}) = |\phi(\textbf{p})|^2.
\end{equation}
This is the first success. The model can reproduce perfectly any single
particle momentum distribution. On the other hand

\begin{equation}\label{}
  C_2(\textbf{p}_1,\textbf{p}_2) = 1 + |\langle e^{iqx_s}\rangle|^2,
\end{equation}
where the averaging is over the distribution of sources $\rho(x_s,p_s)$. Here
functions $\phi(\textbf{p})$ cancel. One way of calculating the distribution
$\rho(p_s,x_s)$ is to start with some initial distribution and then to
propagate it finding from some classical equations, e.g. Newton's or
Boltzmann's, the functions $x_s(t)$ and $p_s(t)$.

Another variant, known as the method of covariant currents \cite{GKW}, is to
put

\begin{equation}\label{}
  \psi_{x_sp_s}(\textbf{p}) = e^{ipx_s}j\left(\frac{p_sp}{m_s}\right),
\end{equation}
where $m_s$ is the mass of the source usually put equal to the particle mass.
In the rest frame of its source each current reduces to the same function
$j(p_0)$. In this approach the assumption of Kopylov and Podgoretskii that each
source yields particles with the same momentum distribution in some common
frame, e.g. in the center-of-mass frame, is replaced by the more plausible
assumption that the momentum distribution for particles from any source looks
the same in the rest frame of this source.

\section{Wigner functions and their generalizations}

In statistical physics there is a well-known method of including simultaneously
positions and momenta.  One uses the Wigner function related to the density
matrix in the momentum representation by the formula \cite{PRA}, \cite{BIK}

\begin{equation}\label{wigfun}
  W(\textbf{X},\textbf{K}) = \int \frac{d^3q}{(2\pi)^3} \rho(\textbf{K} + \frac{1}{2}\textbf{q},\textbf{K} -
  \frac{1}{2}\textbf{q})e^{i\textbf{qX}},
\end{equation}
where

\begin{equation}\label{}
  K = \frac{1}{2}(p + p');\qquad X = \frac{1}{2}(x + x').
\end{equation}
Note that, for further use, $K$ and $X$ are defined as four-vectors, but in the
definition of the Wigner function only three-vectors and a fixed time argument
(not written explicitly) appear. The Fourier transformation can be inverted, so
that there is a one to one relation between the density matrix and the
corresponding Wigner function. The introduction of the Wigner function solves
the problem when the production of all the particles is simultaneous at some
time $t$. When the particles are produced in a time interval $[t_1,t_2]$, but
so that the particles produced at different instants of time do not interfere,
one can overage over time, with suitable weights, both sides of the equation.
The time averaged density matrix, which one could hope to measure, is related
to the time averaged Wigner function, but the important information about the
time distribution of the production process is lost. When the particles
produced at different instants of time interfere, even more information is
lost. It is possible to define objects related by Fourier transformations to
the various components contributing to the density matrix, but from the point
of view of interpretation they are very different from Wigner functions
\cite{ZAL1}.

Formally, one can write \cite{PRA}

\begin{equation}\label{emifun}
  \rho(\textbf{p};\textbf{p}') = \int d^4X e^{iqX}S(X,K),
\end{equation}
where function $S(X,K)$ is known as the emission function. One of the
difficulties is that the four-dimensional Fourier transformations cannot be
inverted. The reason is that for fixed $K$, from $Kq = (p^2 - {p'}^2)/2 = 0$
one finds

\begin{equation}\label{}
  q_0 = \frac{\textbf{Kq}}{K_0},
\end{equation}
while in order to invert the Fourier transformation one needs
$\rho(\textbf{p};\textbf{p}')$, at given $K$, for all values of $\textbf{q}$
and $q_0$. In fact, this difficulty is general and appears as soon as we
introduce time-dependent sources. Thus, there is an infinity of emission
functions $S(X,K)$ which yield the same density matrix
$\rho(\textbf{p};\textbf{p}')$.

For instance one could put

\begin{equation}\label{}
  S(X,K) = \delta(X_0 - t)e^{-iq_0X_0}W(\textbf{X},\textbf{K};t).
\end{equation}
This formula is correct in the sense that, as easily seen using the inverse of
transformation (\ref{wigfun}), it yields the correct density matrix. It is,
however, completely useless, because guessing this Wigner function is just as
hard as guessing the density matrix in the momentum representation. Actually,
this formula has a simple physical interpretation. Let us choose a moment of
time $t_0$, when all the final hadrons are already present and do not interact
any more. From this time on the density matrix for each hadron is

\begin{equation}\label{}
  \rho(\textbf{p},\textbf{p}';t) = e^{-iq_0(t - t_0)}\rho(\textbf{p},\textbf{p}';t_0).
\end{equation}
Thus, it is enough to know $\rho(\textbf{p},\textbf{p}';t_0)$ to predict all
that happens later. What had happened earlier is irrelevant for this
prediction. One can just as well assume that all the hadrons got created at $t
= t_0$. Physically this is a stupid assumption, but formally it is good enough
to predict the states of the system at later times.

One could ask what is the "reasonable" emission function corresponding to the
realistic picture of particle emission. A formula for this function has been
proposed by Shuryak \cite{SHU}. Let us consider first the source labelled $i$.
If it created particles in the pure state $A_i(\textbf{p})$, the density matrix
for the particles produced from this source would be
$A_i(\textbf{p})A_i^*(\textbf{p}')$. Averaging over the parameters of the
source, one obtains the density matrix for the particles produced by source $i$
without the assumption that the state is pure. Summing over the sources one
obtains the overall single particle density matrix

\begin{equation}\label{}
  \rho(\textbf{p},\textbf{p}') = \sum_i\langle A_i(\textbf{p})A_i^*(\textbf{p}')\rangle.
\end{equation}
Expressing the amplitudes $A_i(\textbf{p})$ in terms of their Fourier
transforms $J_i(x)$ one can write for source $i$

\begin{equation}\label{}
\langle A_i(\textbf{p})A_i^*(\textbf{p}')\rangle  = \int d^4X \int d^4Y e^{iqX
+ iKY}\langle J_i(X + \frac{1}{2}Y)J_i^*(X - \frac{1}{2}Y)\rangle,
\end{equation}
where $Y = x - x'$. Summing over $i$ and introducing the notation

\begin{equation}\label{}
\langle J(X + \frac{1}{2}Y)J^*(X - \frac{1}{2}Y)\rangle  = \sum_i\langle J_i(X
+ \frac{1}{2}Y)J_i^*(X - \frac{1}{2}Y)\rangle
\end{equation}
one finds

\begin{equation}\label{}
\rho(\textbf{p},\textbf{p}') = \int d^4X \int d^4Y e^{iqX + iKY}\langle J(X +
\frac{1}{2}Y)J^*(X - \frac{1}{2}Y)\rangle
\end{equation}
Comparing this formula with the (ambiguous!) definition of the emission
function (\ref{emifun}) one finds

\begin{equation}\label{}
  S(X,Y) = \int d^4Y e^{iKY}\langle J(X +
\frac{1}{2}Y)J^*(X - \frac{1}{2}Y)\rangle.
\end{equation}

The strategy is to guess the emission function, with some free parameters,
using all the available information about the sources. Then, one calculates the
density matrix and the correlation function $C_2$ in order to fix the
parameters by comparison with experiment. The weak point is that, since there
is an infinite variety of different emission functions which all give the same
density matrix, a good fit to the data does not necessarily mean that the
models being used, and consequently the parameters obtained, make sense.

\section{Conclusions}

The single particle momentum distribution gives the diagonal element of the
single particle density matrix. Under some simplifying assumptions the
correlations due to quantum statistics give information about the out of
diagonal elements of this matrix. Even if we could find from experiment the
complete single particle density matrix in the momentum representation, which
is not quite the case, this is not enough to find unambiguously the space-time
distribution of the sources. On the other hand, given a good model, it is
possible to fit the experimental data related to the density matrix and thus
find some parameters of the model.

The classical full description: for each particle find its momentum and the
point in space-time  where  the particle has been produced, cannot be achieved
respecting the rules of quantum mechanics. Two popular compromises are: use the
positions and momenta of the sources instead of the positions and momenta of
the particles or use $X$ and $\textbf{K}$ instead of $x$ and $\textbf{p}$.
Given the general framework one needs some phenomenological assumptions to fill
it. An example, admittedly not a very realistic one, is to assume that all the
particles are produced simultaneously from a Gaussian distribution of point
sources. In practice, of course, much more sophisticated phenomenology is being
used.

For quantitative work one must include a number of complicating features, which
have been ignored here for the sake of clarity and of economy of time. We give
below a partial list. More information can be found in the reviews \cite{WIH},
\cite{CSO}.
\begin{itemize}
  \item Many body symmetrization.
  \item Final state interactions.
  \item Production of resonances.
  \item Momentum-position correlations.
  \item Evolution of the interaction region before and during
  hadronization.
\end{itemize}

\end{document}